\documentclass[useAMS,usenatbib]{mn2e}
\usepackage{graphicx}
\usepackage{amssymb}
\usepackage{amsmath}
\title[Coupling of stellar disc with molecular
  torus]{The coupling of a young stellar disc with
  the molecular torus in the Galactic centre}
\author[J.\,Haas, L.\,\v{S}ubr, and P.\,Kroupa]{J.\,Haas$^{1}$\thanks{E-mail:
  haas@sirrah.troja.mff.cuni.cz}, L.\,\v{S}ubr$^{1,2}$, and
  P.\,Kroupa$^{3}$\\
  $^{1}$Astronomical Institute, Faculty of Mathematics and Physics,
  Charles University, V Hole\v{s}ovi\v{c}k\'{a}ch
  2, 18000 Praha, Czech Republic\\
  $^{2}$Astronomical Institute, Academy of Sciences, Bo\v{c}n\'{i} II,
  14131 Praha, Czech Republic\\
  $^{3}$Argelander Institute for Astronomy (AIfA), University of Bonn,
  Auf dem H\"{u}gel 71, 53121 Bonn, Germany}
\begin{document}
\date{Accepted ---. Received ---; in original form ---}
\pagerange{\pageref{firstpage}--\pageref{lastpage}} \pubyear{2010}
\maketitle
\label{firstpage}
%
%
\begin{abstract}
The Galactic Centre hosts, according to observations, a number of
early-type stars.
About one half of those which are orbiting
the central supermassive black hole
on orbits with projected radii $\gtrsim0.03~\mathrm{pc}$
form a coherently rotating disc.
Observations further reveal
a massive gaseous torus and a significant population of late-type stars.
In this paper, we investigate, by means of numerical $N$-body computations,
the orbital evolution of the stellar disc,
which we consider to be initially thin. We include the
gravitational influence of both the torus and the late-type stars,
as well as the self-gravity of the disc.
Our results show that, for
a significant set of system parameters, the evolution of the disc
leads, within the lifetime of the early-type stars,
to a configuration compatible with the observations. In particular,
the disc naturally reaches a specific -- perpendicular -- orientation
with respect
to the torus, which is indeed the configuration observed in the Galactic
Centre.
We, therefore, suggest
that all the early-type stars
may have been born within a single gaseous disc.
\end{abstract}
\begin{keywords}
stellar dynamics -- Galaxy: nucleus.
\end{keywords}
%
%
\section{Introduction}
Over the past two decades, nearly 200 early-type stars have been
revealed in the innermost parsec of our Galaxy
\citep[see Genzel et al. 2010 for the most recent
review;][]{b2,b10,b11,b12,b18,b3,b4}.
Observations suggest that these stars are
orbiting a highly concentrated mass, which is associated with the compact
radio source Sgr~A*. It is widely accepted that this source
is powered by a supermassive black hole (SMBH). Its mass and distance from
the Sun are estimated to be
approximately $4\times10^{6}\,M_{\odot}$ and $8~\mathrm{kpc}$, respectively
\citep{b11,b9,b13,b25,b21}.

According to the most recent observations of \citet{b3,b4},
the majority (136)
of the early-type stars observed in the Sgr~A* region are located
at projected distance $0.03~\mathrm{pc}\lesssim r \lesssim0.5~\mathrm{pc}$
from the SMBH. Roughly one half of these stars appear to
form a coherently rotating disc-like structure, the
so-called clockwise system \citep[CWS; discovered by][]{b16}.
The remaining stars are randomly scattered off the CWS plane. Nevertheless,
some authors report the
existence of a second coherent structure at a large angle with respect
to the CWS -- the counterclockwise
system \citep[CCWS; first mentioned by][]{b10}.
Given these two discs, a significant
number of the
early-type stars are still not belonging to either of the structures.

Observations have further established that all the early-type stars
between $0.03~\mathrm{pc}$ and $0.5~\mathrm{pc}$ from the SMBH
are either Wolf-Rayett stars or O- or early B-stars (WR/OB stars)
\citep{b3,b4}. Evolutionary phases of individual stars indicate that all of
them have been formed $6\pm2~\mathrm{Myr}$ ago within a short period
of time, probably not exceeding $2~\mathrm{Myr}$
\citep{b18}.
The presence of such stars so close to
a SMBH is rather suprising. In particular, the tidal field of the SMBH
is strong enough
to prevent standard star formation mechanisms. Hence, various hypotheses
have been suggested to explain the origin and
configuration of the WR/OB stars observed in the Sgr~A* region.

In situ fragmentation of a self-gravitating gaseous disc
is probably the currently most widely accepted formation scenario
for the stars of the CWS
\citep{b16,b18}. This process was theoretically predicted
to form stars in
active galactic nuclei around SMBHs of masses
$10^{6}$--$10^{10}\,M_{\odot}$ \citep{b6}. However, as it naturally forms stars
in a single disc-like structure, it
fails to explain
the origin of the stars observed outside the CWS.
Many authors have, therefore, been seeking a mechanism that could have scattered
these outliers from the parent disc plane.

It has been shown by \citet{b7} that two-body relaxation of the parent
disc does not yield the observed large inclinations of the outliers with
respect to the disc plane. According to \citet{b26},
some of them may have
been brought to their
positions by vector resonant relaxation between the disc and the
cluster of late-type
stars, which has also
been reported in the Sgr~A* region \citep{b10,b19,b8}.
However, it is still unclear whether this process can
explain the origin of the stars with line-of-sight angular momenta opposite
to that of the stars within the CWS.
In order to overcome this issue, \citet{b17} have
considered
mutual interaction of two self-gravitating discs at large angles
relative to each other.
Although this mechanism indeed yields the observed configuration of the
WR/OB stars, it needs rather special
initial conditions. In particular, the two
discs must have been formed at
specific angles with respect to each other in order to stand for the CWS and
CCWS. Moreover, as all the WR/OB stars
seem to be coeval \citep{b18},
the two discs must have had started to form stars at
the same time. Since this is not very likely, the need of such special
initial conditions
represents the major drawback of this scenario.
Similar scenarios, such as the interaction of two
gaseous streams \citep{b14}, suffer from the
same problem.

\citet{b20} and \citet{b24} have suggested that all the WR/OB stars
in the Sgr~A* region may have been born in a \emph{single} gaseous disc.
They argue that the stars observed outside the CWS represent the outer
parts of the parent disc,
that have been partially disrupted by the gravity of the
circumnuclear disc (CND). The CND is a clumpy molecular torus, that
is located between $1.6~\mathrm{pc}$ and $2.0~\mathrm{pc}$ from
Sgr~A* \citep{b5}. The total mass of
this structure, which is almost perpendicular to the
CWS \citep{b18},
is estimated to be of the order of $10^6\,M_{\odot}$.
\citet{b20} claim that the gravity of the CND
would cause differential precession of
the individual orbits in the parent stellar disc.
Such a process would force the stars from the outer parts of the disc to
leave the disc plane while
the inner parts of the disc would remain untouched. This core would
be identified as the CWS today.

In this paper, we further investigate the hypothesis of \citet{b20}.
In particular, we
include the self-gravity of the parent stellar disc, and follow its
orbital evolution
in a predefined external potential
by means of numerical $N$-body computations.
In addition to the SMBH and the CND, the external potential includes
the gravity of the cluster
of late-type stars. Even though its density
profile is still unclear, its potential
may be considered, in the first approximation,
to be spherically symmetric, and centred
on the SMBH.

In the following section, we briefly introduce the problem
of stellar dynamics in a perturbed Keplerian potential
and describe our model. The results of our calculations are presented
in Section
\ref{results} and discussed in Section \ref{discussion}. We conclude our work
in Section \ref{conclusions}.
%
%
\section{Orbital evolution of a stellar disc in an external potential}
The gravitational potential in the vicinity of the CWS induced by the SMBH
is, to a very high accuracy, Keplerian. Hence,
it is useful to describe the individual stellar orbits
by means of Keplerian orbital elements: semi-major axis $a$,
eccentricity $e$,
inclination $i$, longitude of the ascending node $\Omega$, and argument of
pericentre $\omega$. The orbital evolution of the disc can then be
investigated
by following the evolution of these elements.
If the only component of the external potential were Keplerian gravity of the
SMBH, and if the
stars in the disc were treated as test particles, the orbital elements
would remain constant in time.
On the other hand, with any additional potentials included,
some of the elements
may undergo complex secular evolution.

\citet{b20} have investigated the influence of the CND
upon the orbital
evolution of the disc. They have considered the disc to be
surrounded by the cluster of late-type stars, and the stars in the
disc to be test particles.
They claim that the
evolution of individual stellar orbits in the disc
is dominated by two processes:
(i) precession of the ascending node,
and (ii) periodic oscillations of eccentricity and inclination
(Kozai oscillations; independently theoretically predicted by
Kozai 1962 and Lidov 1962).
These oscillations are diminishing
with increasing mass of the cluster of late-type stars.
They become fully negligible when the
mass, $M_{\mathrm{c}}$, of the cluster
within the radius $R_{\mathrm{CND}}$ of the CND fulfils the
approximate condition
$M_{\mathrm{c}}\gtrsim0.1\,M_{\mathrm{CND}}$, where $M_{\mathrm{CND}}$
stands for the mass of the CND.
In that case,
the first time derivative of $\Omega$ becomes constant and can
be writen as
\begin{equation}
  \frac{\mathrm{d}\Omega}{\mathrm{d}t}=-\frac{3}{4}
  \frac{\cos{i}}{T_{\mathrm{K}}}
  \frac{1+\frac{3}{2}e^2}{\sqrt{1-e^2}}
\label{rate}
\end{equation}
with
\begin{equation}
  T_{\mathrm{K}}\equiv\frac{M_{\mathrm{\bullet}}}{M_{\mathrm{CND}}}
  \frac{R_{\mathrm{CND}}^3}{a\sqrt{GM_{\bullet}a}}\,,
\end{equation}
where $G$ denotes the gravitational constant and $M_{\bullet}$ represents
the mass of the SMBH.
According to this formula, the rate of precession strongly depends upon the
semi-major axis of the orbit. Hence, the outer parts of the disc are more
affected by the precession than the inner parts and, therefore,
the disc becomes warped or, eventually, completely disrupted.

In formula (\ref{rate})
and further on, we assume all angles to be measured in
the frame where the CND lies in the $xy$-plane. Consequently, inclination
of the orbit $i$
is the angle between the symmetry axis of the CND and angular momentum of the
star.
\begin{figure*}
  \includegraphics[width=\textwidth]{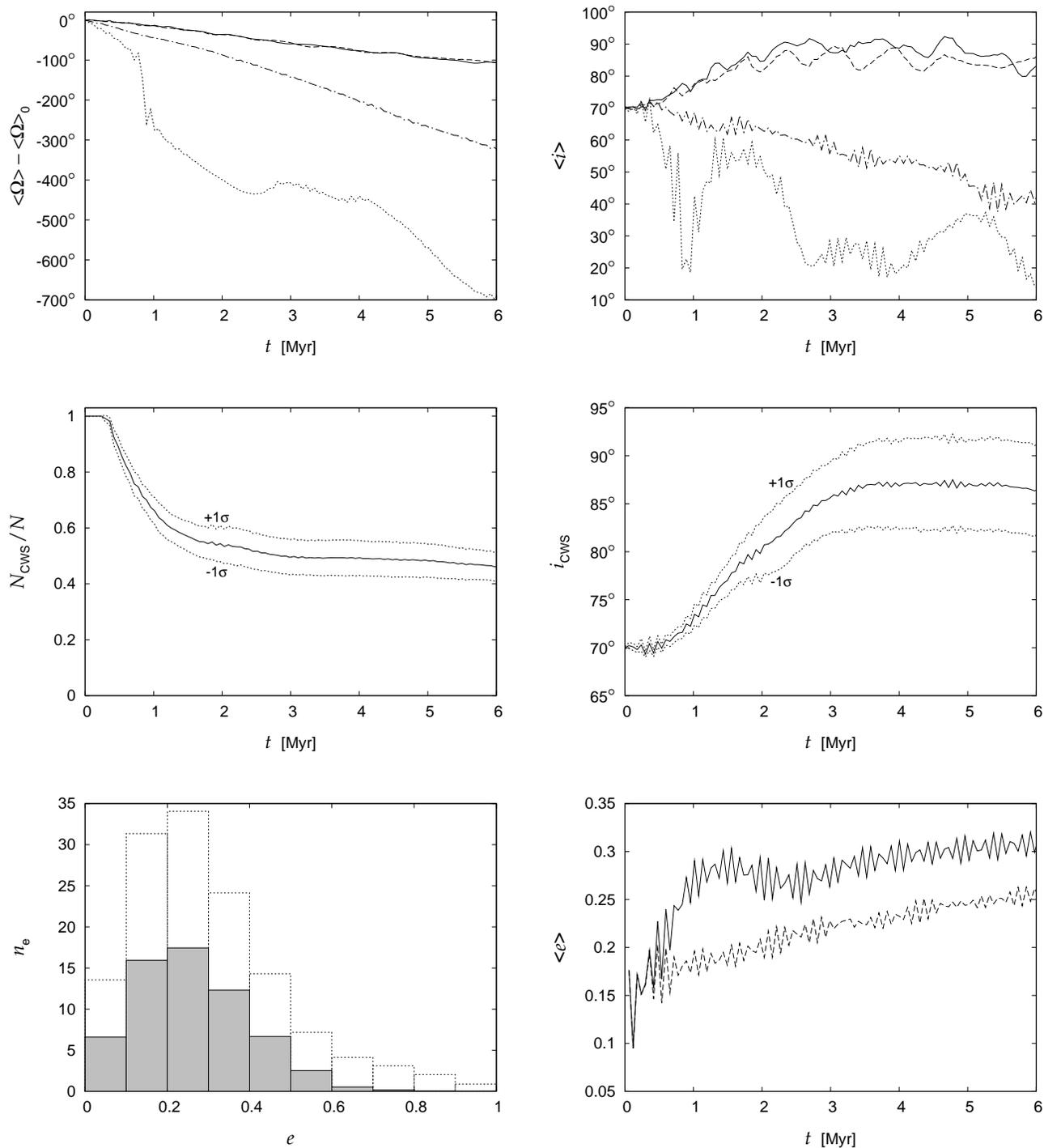}
  \caption{Results for the canonical model (see Table\,\ref{canonical}
  for the corresponding parameters). Only properties of the stars with
  $m\geq12\,M_{\odot}$ are displayed.
  The dotted lines in the middle panels denote
  standard deviation for the set of 120 included realisations.
  Top: Evolution of the mean
  value of $\Omega$ (left) and $i$ (right) within different
  parts of the disc for one of the realisations of this model. The
  solid line describes the group of the innermost stars,
  followed by the dashed,
  dot-dashed and dotted line, which correspond to successive outer groups.
  Middle-left: Number of stars within the CWS (i.\,e. with
  angular momentum deviating from the mean angular
  momentum of the CWS by less than $20^{\circ}$).
  Middle-right: Inclination of the CWS with respect to the CND.
  Bottom-left: Eccentricity distribution for the stars
  after $6~\mathrm{Myr}$ of orbital evolution. The empty boxes
  denote distribution for all the stars in the
  young stellar system while the
  grey ones represent only stars within the CWS. Bottom-right: Mean
  eccentricity of all the stars in the
  young stellar system (solid line) and
  within the CWS (dashed line).}
  \label{best}
\end{figure*}

With the gravity of the stars in the disc included, the
orbital elements of individual orbits undergo random variations
due to two-body relaxation.
As a result, the mean value of eccentricity
and inclination
of the orbits increases.
These changes are, by themselves,
too small to have a significant impact on the overall
shape of the disc \citep{b7}. However, according to formula (\ref{rate}),
they accelerate the differential precession of the orbits.
Let us, therefore,
further investigate the differential precession of the orbits within
a self-gravitating disc.
For this purpose, we introduce the model of the
Galactic Centre in the following way.
\begin{enumerate}
  \item{}The SMBH of mass $M_{\bullet}=4\times10^6\,M_{\odot}$ is
    considered to be
    a source of Keplerian potential.
  \item{}The CND is modelled as a single massive particle of mass
    $M_{\mathrm{CND}}$ orbiting the SMBH on a circular orbit
    of radius $R_{\mathrm{CND}}=1.8~\mathrm{pc}$ (see Subsection \ref{CND} for
    the discussion of this approximation).
  \item{}The cluster of late-type stars is represented
    by a smooth power-law
    density profile, $\rho\left(r\right)\propto r^{-\beta}$, and mass
    $M_{\mathrm{c}}$ within the radius $R_{\mathrm{CND}}$.
  \item{}The early-type stars in the disc are treated as
    $N$ gravitating particles,
    $m\in[m_{\mathrm{min}},m_{\mathrm{max}}]$,
    distributed according to a power-law mass function
    $\mathrm{d}N\propto m^{-\alpha}\mathrm{d}m$.
\end{enumerate}

The stellar orbits
in the disc are constructed to be initially geometrically circular.
However, due to the additional spherically symmetric component of the
gravitational potential (cluster of late-type stars),
the osculating eccentricities
do not truthfully describe real curvature of the orbits in space. E.\,g.
the initial osculating
eccentricity
of the outermost orbits in the disc is $\approx0.1$
for $M_{\mathrm{c}}=0.1$ and $\beta=7/4$.
Initial radii of the orbits are, in accord with the observations
\citep{b18,b3,b4},
generated randomly between $0.04~\mathrm{pc}$ and $0.4~\mathrm{pc}$.
Their distribution obeys
$\mathrm{d}N\propto a^{-1}\mathrm{d}a$.
The disc is initially thin with
half-opening angle
$\Delta_0\lesssim5^{\circ}$. The initial inclination
of the disc plane with respect to the CND,
which is defined by the mean angular momentum of the stars in the disc,
is denoted $i_{\mathrm{CWS}}^0$.

We follow the evolution of the latter
system numerically by means of the
$N$-body integration code NBODY6 \citep{b1}. The gravitational
potentials of both the SMBH and the cluster of late-type stars
have been incorporated into the original code as
additional external potentials.
%
%
\section{Results}
\label{results}
The evolution of the stellar disc depends upon the shape of the
gravitational potential, which is determined by the parameters
$M_{\mathrm{CND}}$, $M_{\mathrm{c}}$, $\beta$, $N$,
$m_{\mathrm{min}}$, $m_{\mathrm{max}}$, $\alpha$, $\Delta_0$, and
$i_{\mathrm{CWS}}^0$.
Hence, in order to investigate the evolution
properly, it is necessary to cover all the reasonable
values of these parameters.
On the other hand,
in order to demonstrate the results of our
calculations, it is useful to define a ``canonical'' model
with the parameters set to the values listed in Table \ref{canonical}.

The observations suggest that all the WR/OB stars have mass
$m\gtrsim12\,M_{\odot}$ \citep{b18,b3,b4}.
However, since it is likely that a number of
undetected less massive early-type stars exist in the Sgr~A* region,
we consider $m\in[4\,M_{\odot},120\,M_{\odot}]$
in the ``canonical'' model.
Nevertheless, for a more
convenient comparison with the currently available observational data,
we display properties of only a subset
of the stars with mass $m\geq12\,M_{\odot}$ in figures.

Due to the stochastic nature of the studied system, the results
should be averaged over a number of realisations
with identical values of the model parameters
in order to distinguish general trends from random fluctuations.
For this purpose, we first considered 120 realisations of the
``canonical'' model.
It has, however, turned out that the results
become statistically relevant already for 12 realisations.
Hence, we consider only 12 realisations of all
the other models discussed in this paper
in order to shorten the necessary computational time
($\approx5$ hours on $3~\mathrm{GHz}$ CPU per run; $\approx2000$ runs in total).

Since we attempt to explain the configuration
of a specific observed system, every single realisation represents a possible
course of its evolution. We thus show
the standard deviation for some of the
key quantities for a more thorough description of the set of
possible evolutions.
\begin{table}
\caption{Parameters of the ``canonical'' model (see Fig.\,\ref{best}
for the corresponding results).}
\begin{center}
\begin{tabular}{|r@{\ =\ }l|r@{\ =\ }l|r@{\ =\ }l|}
\hline
$M_{\bullet}$&
$4\times10^6\,M_{\odot}$&
$M_{\mathrm{CND}}$&
$0.3\,M_{\bullet}$&
$M_{\mathrm{c}}$&
$0.03\,M_{\bullet}$\\
$m$&
$4$--$120\,M_{\odot}$&
$\alpha$&
$1$&
$\beta$&
$7/4$\\
$N$&
$200$&
$i_{\mathrm{CWS}}^0$&
$70^{\circ}$&
$\Delta_0$&
$2.5^{\circ}$\\[4pt]
\multicolumn{6}{c}{Total
  mass of the young stellar disc $\approx6.8\times10^3\,M_{\odot}$}\\
\multicolumn{6}{c}{$\left(\gtrsim90\,\%\right.$ by
  $\approx140$ stars with $\left.m\geq12\,M_{\odot}\right)$}\\
\hline
\end{tabular}
\end{center}
\label{canonical}
\end{table}

The results for the canonical model are shown in Fig.\,\ref{best}.
The top-left panel
demontrates the differential precession of the orbits in the disc.
It shows the evolution of the mean value of
$\Omega$ within different groups of stars, which are determined by their
initial distance from the centre.
It turns out that the precession of the ascending node affects
more strongly
the orbits in
the outer parts of the disc (dotted and dot-dashed lines)
than those in the inner parts
(dashed and solid lines). This result
is in accord
with formula (\ref{rate}) and proves the gradual
deformation of the disc.

Our results further show that
the precession of the ascending node
in the outer
parts of the disc is globally accelerated. We attribute this effect,
which becomes significant on longer time scales,
to the evolution of inclination due to two-body relaxation of the disc.
Such an acceleration was not found by
\citet{b20} as they
had neglected the gravity of the stars in the disc.
\begin{figure}
  \includegraphics[width=\columnwidth]{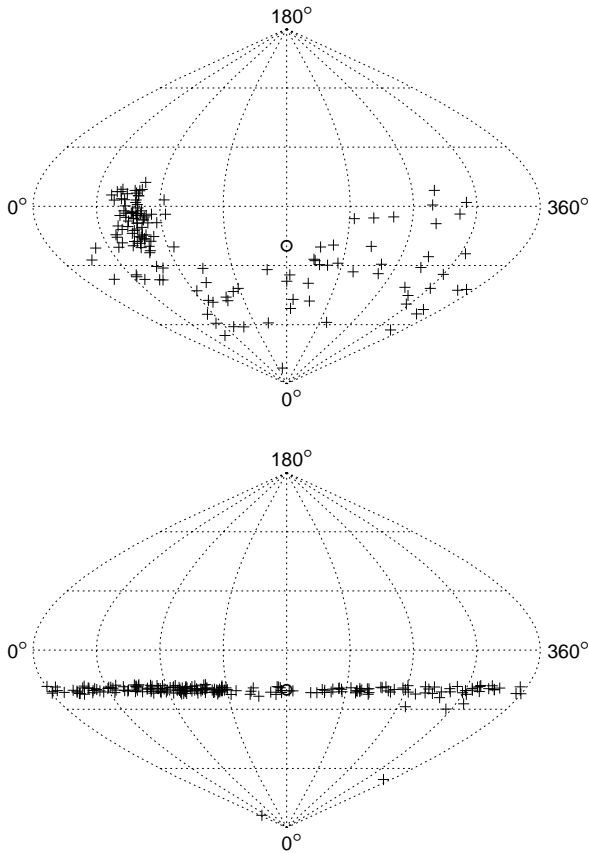}
  \caption{Angular momenta of individual stars in the
  young stellar disc in sinusoidal
  projection after $6~\mathrm{Myr}$ of orbital evolution.
  The initial state is denoted by an empty circle. Latitude on the plots
  corresponds to $i$ while longitude
  is related to $\Omega$. The top panel shows the results for one
  of the realisations of the canonical
  model
  (only stars with $m\geq12\,M_{\odot}$ displayed). For comparison, the bottom
  panel illustrates the situation
  with negligibly small mass of the stars in the disc (single mass,
  $m=0.004\,M_{\odot}$, the other parameters are the same
  as in the canonical model).}
  \label{Omegavsi}
\end{figure}

The sudden drop
of $\langle\Omega\rangle$ on the dotted line in the
top-left panel of Fig.\,\ref{best} is a residue
of Kozai oscillations. Since the
cluster of late-type stars
is, in our canonical model, not massive enough to
suppress the oscillations of $e$ and $i$ entirely, the
first time derivative of $\Omega$ also varies on the timescale of
$T_{\mathrm{K}}$.

The evolution of the mean inclination $\langle i\rangle$
with respect to the CND within different parts of the disc
is shown in the top-right panel of Fig.\,\ref{best}.
We can see that the inclination of the outer parts of the disc
is decreasing
(dotted and dot-dashed lines), while it grows and
saturates at $\approx90^{\circ}$ in the inner parts
(dashed and solid lines).
We further find that
the evolution of both $\Omega$ and $i$
is similar for
all the orbits in the inner parts of the disc.
Hence, the core
of the disc remains rather undisturbed and coherently changes its
orientation towards perpendicular with respect to the CND.
This effect can be seen in the
top panel of Fig.\,\ref{Omegavsi}, which shows
the directions of angular
momenta of the individual stars in the disc
after $6~\mathrm{Myr}$ of orbital evolution
for one of the realisations of the canonical model
(the initial state is denoted by an empty circle).
Our results prove that the compact group at inclination $\approx90^{\circ}$
is formed by the stars from the inner parts of the disc,
while the remaining scattered stars
represent the entirely dismembered outer parts.
Hence, we see that the dynamical evolution
of the initially thin stellar disc leads to a configuration similar
to that observed in the Sgr~A* region
\citep[see][]{b18,b3,b4}. In particular, the core of
the disc can be identified with the CWS observed today and, at the
same time, the stars from the dissolved outer parts can stand for the
WR/OB stars found outside the CWS.

In order to compare our results with the observations more thoroughly,
we further define CWS within our model
in the following iterative way.
As the zeroth step, the CWS is
considered to be formed by a fixed number of the
innermost stars from the initial disc. In the next step, we exclude from
the CWS
all the stars whose angular momenta deviate from the mean angular momentum
of the CWS by more than $20^{\circ}$. On the other hand, the stars initially
from outside the CWS, which do not fulfil the latter condition, are
included into the CWS. Then, we recalculate the mean angular
momentum of the CWS and repeat the whole procedure iterativelly until
there are no changes of the CWS in between two subsequent steps.

Observations suggest
\citep{b18,b3,b4} that roughly one half of the WR/OB stars are members
of the observed CWS. We thus follow within our calculations
the relative
number of stars, $N_{\mathrm{CWS}}/N$, which belong to the CWS.
As can be seen in the middle-left panel of Fig.\,\ref{best},
this number reaches, within our model, the value of
$\approx0.5$ at $t=6~\mathrm{Myr}$.

As computed by \citet{b18}, the normal vectors
of the observed CWS and the CND fulfil
$\boldsymbol{n}_{\mathrm{CWS}}\cdot\boldsymbol{n}_{\mathrm{CND}}=0.01$,
which corresponds to the mutual angle of $89.4^{\circ}$. In order
to confront this feature,
we investigate the evolution of the inclination $i_{\mathrm{CWS}}$
of the CWS with respect to the CND. Our computations show that
$i_{\mathrm{CWS}}\approx90^{\circ}$ at $t=6~\mathrm{Myr}$
(see the middle-right
panel of Fig.\,\ref{best}), which is
in a remarkable agreement with the observational data.

Finally, we investigate the eccentricity
distribution $n_{\mathrm{e}}$ within the CWS and in the whole
young stellar system
after $6~\mathrm{Myr}$ of orbital evolution.
The corresponding histograms
in the bottom-left panel
of Fig.\,\ref{best} show that in both cases
a substantial fraction of the orbits
have, in accord with the observations, moderate eccentricities.
The mean eccentricity
of the stars within the CWS
(see the dotted line in the bottom-right panel of Fig.\,\ref{best})
is then $\approx0.25$, which is
somewhat lower than
the value $0.36\pm0.06$ recently reported by \citet{b3}.
However, at the current level of accuracy, the observations do not
provide sufficient information for a reliable determination of the
orbital eccentricity for a significant number of the WR/OB stars.
Hence, the eccentricity
criterion should be considered only as supplemental.
\begin{figure*}
  \includegraphics[width=\textwidth]{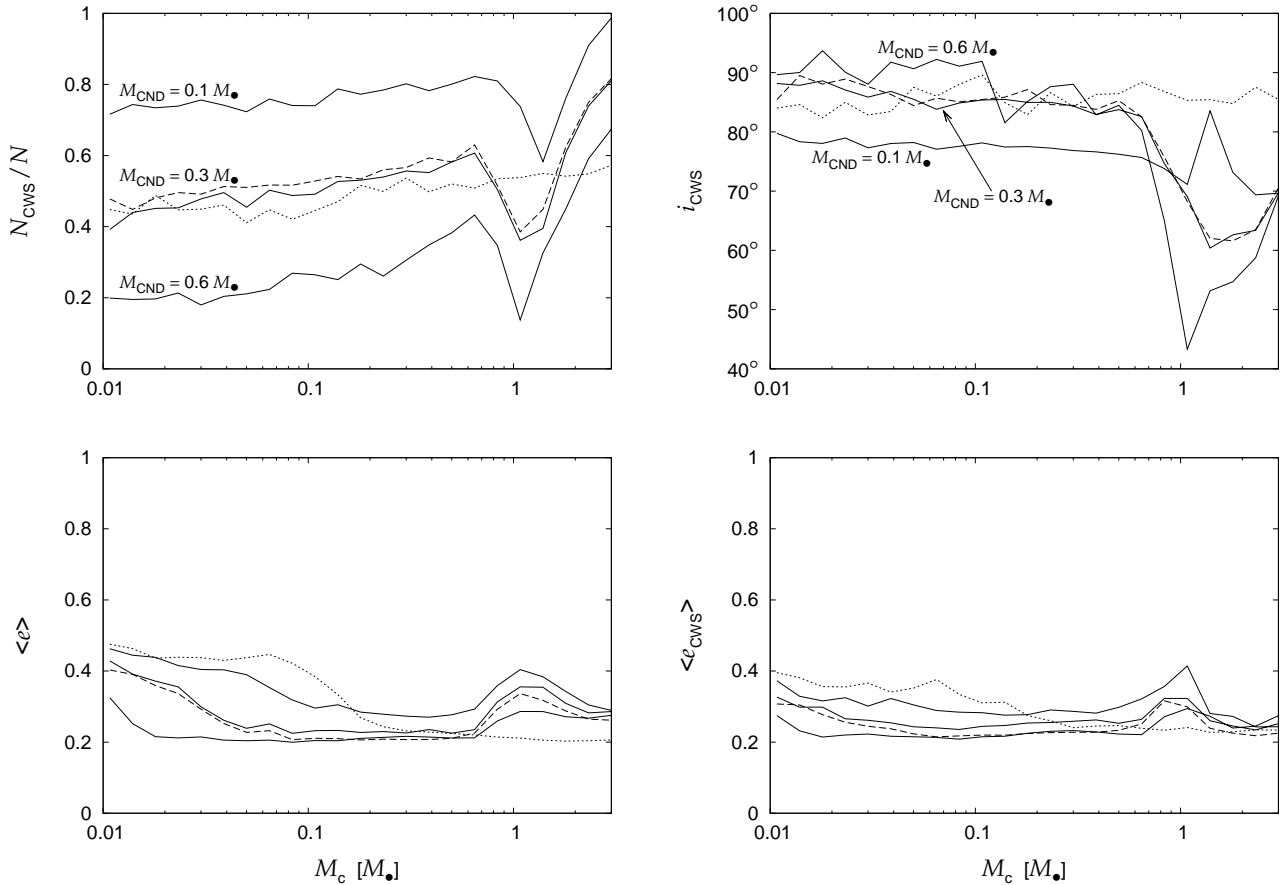}
  \caption{Number of stars within the CWS (top-left),
  inclination of the CWS with respect to the CND (top-right) and
  mean eccentricity of all the stars in the
  young stellar system (bottom-left) and
  within the CWS (bottom-right) at $t=6~\mathrm{Myr}$
  for various values of $M_{\mathrm{CND}}$ and $M_{\mathrm{c}}$.
  All results are averaged over 12 realisations.
  Solid
  lines: $\beta=7/4$, $N=200$,
  $m\in[4\,M_{\odot},120\,M_{\odot}]$ (only properties of stars with
  $m\geq12\,M_{\odot}$ displayed), $\alpha=1$. Dashed line:
  $M_{\mathrm{CND}}=0.3\,M_{\bullet}$,
  $\beta=1/2$, $N=200$, $m\in[4\,M_{\odot},120\,M_{\odot}]$
  (only properties of stars with
  $m\geq12\,M_{\odot}$ displayed), $\alpha=1$.
  Dotted line:
  $M_{\mathrm{CND}}=0.3\,M_{\bullet}$,
  $\beta=7/4$, $N=136$, single mass, $m=50\,M_{\odot}$.
  Common parameters for all models are:
  $M_{\bullet}=4\times10^{6}\,M_{\odot}$,
  $\Delta_0=2.5^{\circ}$,
  $i_{\mathrm{CWS}}^0=70^{\circ}$.}
  \label{discuss}
\end{figure*}
%
%
\section{Discussion}
\label{discussion}
As we have shown in the previous section, all the
WR/OB stars may have been formed within a single gaseous disc and, subsequently,
brought to their present location
by the combined effects of two-body
relaxation and differential precession.
In the following we establish a set of parameters
for which the evolution of the
young stellar system leads to a configuration compatible
with the current observational data.
For this purpose, we follow the evolution of the system for various values
of the model parameters. Within the results,
we then concentrate
on $N_{\mathrm{CWS}}/N$, $i_{\mathrm{CWS}}$,
$\langle e_{\mathrm{CWS}}\rangle$ and
$\langle e\rangle$
and confront their values at $t=6~\mathrm{Myr}$
with the observations.
%
%
\subsection{System parameters compatible with the observations}
To begin with, we
follow the evolution of the young stellar system for various values
of $M_{\mathrm{CND}}$ and $M_{\mathrm{c}}$ with the other parameters
fixed to their canonical values (see Table \ref{canonical}).
The results indicate that
the strongest constraints on the possible values of $M_{\mathrm{CND}}$
and $M_{\mathrm{c}}$
come from $N_{\mathrm{CWS}}/N$ and $i_{\mathrm{CWS}}$.
Their values for
$M_{\mathrm{c}}\in[0.01\,M_{\bullet},3\,M_{\bullet}]$
are depicted
by the solid lines in the top panels of Fig.\,\ref{discuss}, whereas
$M_{\mathrm{CND}}$ remains constant along each of the lines
and is
set to either $0.1\,M_{\bullet}$ or $0.3\,M_{\bullet}$ or $0.6\,M_{\bullet}$.
According to these results, the evolution of the young stellar disc leads to
values of $N_{\mathrm{CWS}}/N$ which accommodate
the observational constraints, if
$0.1\,M_{\bullet}\lesssim M_{\mathrm{CND}}\lesssim0.3\,M_{\bullet}$ and
$0.01\,M_{\bullet}\lesssim M_{\mathrm{c}}\lesssim2M_{\bullet}$. However,
the upper limit for the mass of the cluster
of late-type stars has to be reduced to $M_{\mathrm{c}}\lesssim M_{\bullet}$
since larger values do not lead to the observed
$i_{\mathrm{CWS}}\approx90^{\circ}$.
Both observational criteria are, therefore,
fulfilled if $0.1\,M_{\bullet}\lesssim M_{\mathrm{CND}}\lesssim0.3\,M_{\bullet}$
and $0.01\,M_{\bullet}\lesssim M_{\mathrm{c}}\lesssim M_{\bullet}$.
Moreover,
the results of
our computations with $M_{\mathrm{c}}=0$ show that even in this case,
the evolution of the
young stellar system leads
to a configuration which matches the observational data
(due to logaritmic scale in Fig.\,\ref{discuss}, the corresponding values
are not displayed).
Hence, we find the final intervals
$0.1\,M_{\bullet}\lesssim M_{\mathrm{CND}}\lesssim0.3\,M_{\bullet}$
and $0\leq M_{\mathrm{c}}\lesssim M_{\bullet}$.
If we substitute $M_{\bullet}=4\times10^{6}\,M_{\odot}$,
the intervals transform to $4\times10^5\,M_{\odot}\lesssim
M_{\mathrm{CND}}\lesssim1.2\times10^6\,M_{\odot}$ and
$0\leq M_{\mathrm{c}}\lesssim4\times10^{6}\,M_{\odot}$.

The intervals for allowed $M_{\mathrm{CND}}$ and $M_{\mathrm{c}}$
are not affected if we evaluate the eccentricity
criterion. As can be seen in the bottom panels of Fig.\,\ref{discuss}
(solid lines),
all the considered values of $M_{\mathrm{CND}}$ and $M_{\mathrm{c}}$
lead to similar values of both $\langle e_{\mathrm{CWS}}\rangle$ and
$\langle e\rangle$, which
satisfy the observational constraints. Nevertheless, our results
show that the orbital eccentricities in the young stellar system
reach slightly higher values for lower $M_{\mathrm{c}}$. We attribute
this effect to
Kozai oscillations, which are less suppressed by the cluster
of late-type stars. Furthermore, our results suggest
that also for $M_{\mathrm{c}}\approx M_{\bullet}$, all the
orbits in the young stellar system gain somewhat larger eccentricities,
regardless the mass of the CND. Around the same value,
$i_{\mathrm{CWS}}$ appears to be more sensitive upon the variations of
$M_{\mathrm{c}}$ and
$N_{\mathrm{CWS}}/N$ reaches its minima
(see the solid lines in the top panels of Fig.\,\ref{discuss}).
Hence, it seems that all of these effects are somehow connected with a stronger
influence of the CND on the dynamical evolution of the
young stellar disc. However,
at this point, we can not provide any explanation of this effect.

In order to investigate whether the suggested intervals for
$M_{\mathrm{CND}}$ and $M_{\mathrm{c}}$ depend upon the density profile
of the cluster of late-type stars, we model the evolution of the
young stellar system also
for $\beta=1/2$ and $M_{\mathrm{c}}\in[0.01\,M_{\bullet},3\,M_{\bullet}]$.
The other parameters remain at their canonical values
(see Table \ref{canonical}).
The dotted line in Fig.\,\ref{discuss} proves
that
$N_{\mathrm{CWS}}/N$, as well as $i_{\mathrm{CWS}}$ and both
$\langle e_{\mathrm{CWS}}\rangle$ and
$\langle e\rangle$,
reach the same values as in the case
with $\beta=7/4$, except for the neighbourhood of the point
$M_{\mathrm{c}}\approx M_{\bullet}$. The
absence of the resonant effects observed in this case
can be interpreted as a consequence of
the different mass of the cluster of late-type stars enclosed
within the young stellar disc, due to the different value of $\beta$.
However, since the value $M_{\mathrm{c}}\approx M_{\bullet}$ represents only
the approximate upper boundary of the suggested interval for $M_{\mathrm{c}}$,
the latter effect is, for the purpose of this study, rather insignificant.

Similarly, in order to test the dependence of the intervals upon
the mass function of the young stellar disc itself, we perform
a set of computations with the disc treated as a group of 136
single mass stars each with mass $m=50\,M_{\odot}$.
The other parameters
are set to their canonical values.
In this case, the total
mass of the disc is the same as in all the models, which
we have presented so far ($\approx6.8\times10^3\,M_{\odot}$).
As demonstrated by the dashed line in Fig.\,\ref{discuss},
none of the results depend upon the mass
function of the disc if its total mass is preserved.
\begin{figure}
  \includegraphics[width=\columnwidth]{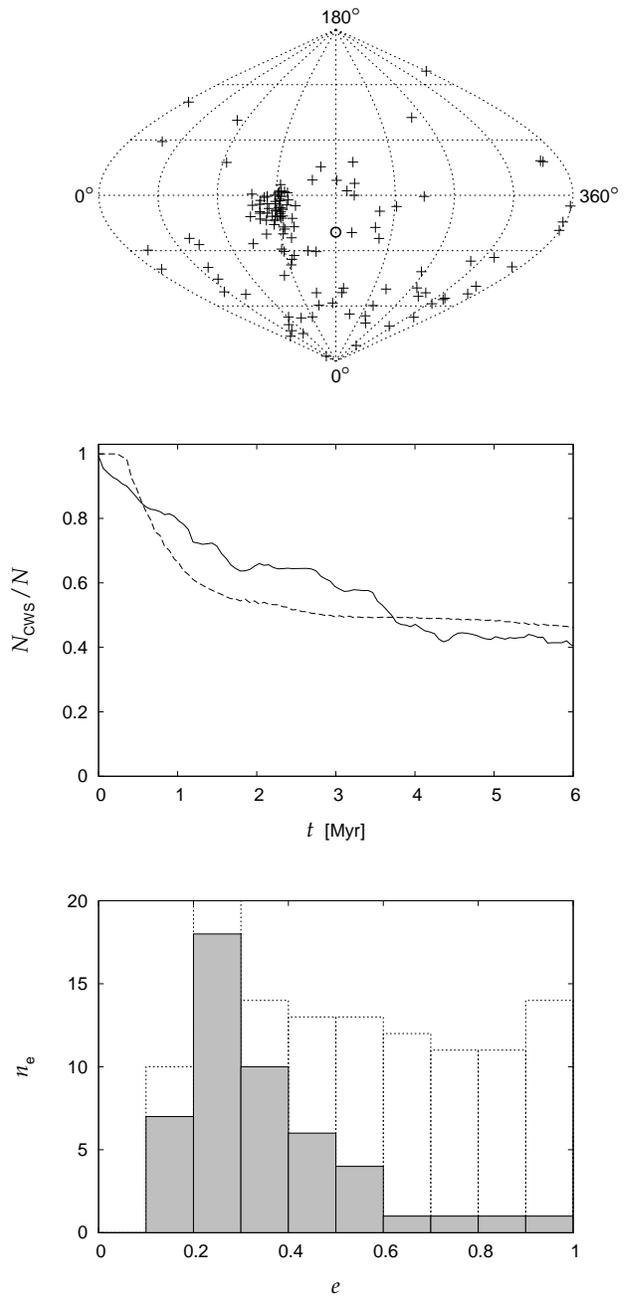}
  \caption{One of the realisations of the model with the CND treated as
  a group of $N_{\mathrm{CND}}=20$ equal-mass particles.
  The other parameters are set to their canonical values (see Table
  \ref{canonical}), except for
  $M_{\mathrm{CND}}=0.1\,M_{\bullet}$ and $M_{\mathrm{c}}=0.1\,M_{\bullet}$.
  Top: Angular momenta of individual stars in the young stellar disc
  at $t=6~\mathrm{Myr}$ (initial state denoted by an empty circle).
  Middle: Number of stars within the CWS (solid line). For comparison,
  we show the results for the canonical model (dashed line).
  Bottom: Eccentricity distribution at $t=6~\mathrm{Myr}$ for all the
  stars in the young stellar
  system (empty boxes) and within the CWS (grey boxes).}
  \label{enhanced}
\end{figure}

Our calculations further show that the evolution of the young stellar disc is
not affected significantly if its total mass is changed
within the range
$\approx10^3$--$10^4\,M_{\odot}$.
On this account,
all the results presented in the previous section would remain
entirely unaffected even if we did not include the undetected
stars with mass $m\in[4\,M_{\odot},12\,M_{\odot}]$ as their
overall mass represents only a small fraction of the total mass of the whole
disc.

On the other hand, decreasing the total mass of the disc by
reducing the mass of the individiual stars $m$ inhibits the combined
effects of two-body relaxation and differential precession. Namely,
we do not observe the evolution of $i_{\mathrm{CWS}}$
if $m$ becomes negligibly small,
i.\,e. if
the stars in the disc can be considered as test particles.
This effect is demonstrated by
the bottom panel of Fig.\,\ref{Omegavsi}, where we set $m = 0.004\,M_{\odot}$.
We can see that except for the few outermost stars, which are still
slightly affected by Kozai oscillations
caused by the CND, the inclination of the orbits
remains constant throughout the disc.
Hence, our results are in accord
with the findings of \citet{b20}.

We have determined the intervals for $M_{\mathrm{CND}}$ and
$M_{\mathrm{c}}$
under assumption $i_{\mathrm{CWS}}^0=70^{\circ}$.
Our results indicate that if $i_{\mathrm{CWS}}^0\gtrsim60^{\circ}$ and
both the $M_{\mathrm{CND}}$ and
$M_{\mathrm{c}}$ fall into the determined intervals, the evolution
of the young stellar system leads
within $6~\mathrm{Myr}$ to a configuration
that agrees with the current observations.
With lower values
of $i_{\mathrm{CWS}}^0$ considered, the CWS is entirely
destroyed
by the differential precession
before it can reach the orientation perpendicular to the CND.
On the other hand, considering $i_{\mathrm{CWS}}^0$ closer to $90^{\circ}$
may increase the allowed intervals for $M_{\mathrm{CND}}$ and
$M_{\mathrm{c}}$.
However, in order to fully understand
the impact of different values of $i_{\mathrm{CWS}}^0$ on the
suggested intervals, a more detailed study would be required.
We will focus on this issue in our future
work.

In this paper, we consider the young stellar disc to be initially thin.
Our calculations show that the course of
its evolution does not depend upon
the value of its initial
half-opening angle if
$\Delta_0\lesssim5^{\circ}$.
%
%
\subsection{Structure of the CND}
\label{CND}
So far, we have modelled the CND as a single massive particle
on a circular orbit around the SMBH. This
approximation has been used instead of
analytical descriptions, e.\,g. infinitesimally thin ring, for numerical
reasons. Namely, due to its simplicity, single-particle approach minimizes
the necessary computational time, and the corresponding perturbing particle can
be implemented into the original NBODY6 code in a trivial way.

The single-particle approximation follows from the
standard averaging technique, which is commonly applied to many problems
of celestial mechanics \citep[see e.\,g.][]{b28}. As a consequence,
the single-particle approximation is equivalent to the model with the CND
treated as an infinitessimally thin ring if the assumptions of the averaging
technique are satisfied. For the studied young stellar system,
these assumptions can be written as the following two conditions
for the orbital period $P_{\mathrm{p}}$ of the massive CND
particle: (i) $P_{\mathrm{p}}$
must be significantly longer than the orbital periods
$P_{\mathrm{d}}^{\mathrm{j}}$ of the
early-type stars in the disc, and (ii) $P_{\mathrm{p}}$
must be significantly shorter than the characteristic period
$P_{\mathrm{c}}$ of the studied phenomena. Since
$P_{\mathrm{c}}\sim10^6~\mathrm{yr}$,
$P_{\mathrm{p}}\sim10^5~\mathrm{yr}$, and
$P_{\mathrm{d}}^{\mathrm{j}}\sim10^{2}$--$10^{4}~\mathrm{yr}$, both
conditions are fulfilled and, therefore, the use of
single-particle approximation is, in our case, well justified.

The real CND is, rather than a ring-like structure, a gaseous torus,
which consists of several
somewhat autonomous clumps. In order to test
whether the evolution of the young stellar disc can be affected by
the clumpiness of the CND,
we further consider the CND to be a group of $N_{\mathrm{CND}}$
equal-mass particles. It turns out that the
CND constructed in this way is unstable with respect to its own gravity.
Consequently,
some of the particles successively migrate towards the SMBH.
These particles can eventually be identified with several
gaseous streams, which
are indeed observed within the radius of the CND
\citep[for one of the most recent studies, see][]{b22}.
The results further show that all the combined effects of
two-body relaxation and differential precession remain present
(see Fig.\,\ref{enhanced} for the case with $N_{\mathrm{CND}}=20$).
Moreover,
the infalling CND particles pose a stronger perturbation for the
young stellar disc. As a result,
the orbits in the disc gain higher eccentricities compared to
models with the CND treated as a single massive particle
on a stable orbit (see
the bottom panel in Fig.\,\ref{enhanced} and the bottom-left
panel in Fig.\,\ref{best}).
Similarly,
the gradual deformation of the young stellar
disc, as well as its eventual destruction, are also accelerated.

Hence,
it appears that the models with the CND treated as a single massive particle
somewhat underestimate its influence upon the dynamical evolution of the young
stellar disc. On the other hand, the perturbative influence of the
infalling parts of the gaseous CND would probably not be as strong as the
impact of infalling point-like particles in the latter model.
A more precise approach to the gas dynamics would thus be required in order
to obtain a more accurate description of the CND.
%
%
\section{Conclusions}
\label{conclusions}
We have modelled the orbital evolution of
a thin self-gravitating stellar disc in a predefined external potential
by means of N-body computations.
In accord with the observations of the Galactic Centre, we have considered
the potential to include the gravity of the SMBH, the CND and the cluster
of late-type stars. The results show that for a significant set of
system parameters, the evolution of the disc leads
to a configuration similar to that observed in the Sgr~A* region
within the lifetime of the WR/OB stars.
In particular, while the outer parts of the disc are entirely dismembered
due to differential precession of the orbits caused by the CND,
the inner parts remain undisturbed forming the CWS.
Due to the influence of the CND,
the CWS tends to change
its orientation towards perpendicular with respect to the CND
for a variety of initial configurations.
Indeed, according to the observations, the CWS and the CND are in such
a specific mutual orientation.
It may thus be plausible
that all the WR/OB stars observed in the Sgr~A*
region may have been born within a single gaseous disc and, subsequently,
brought to their present location by the combined effects of two-body
relaxation and differential precession.

We have further determined the possible values of the mass of the
CND, $M_{\mathrm{CND}}$,
which lead to a configuration compatible with the observational
data.
Based on several observational criteria, we have found the
approximate interval:
$0.1\,M_{\bullet}\lesssim M_{\mathrm{CND}}\lesssim0.3\,M_{\bullet}$.
Although this interval
represents a very strict constraint on the
mass of the CND,
it is in accord with the most recent observational estimate
$M_{\mathrm{CND}}\approx10^6\,M_{\odot}$ found by \citet{b5}.
Moreover, in agreement with the conclusions
of \citet{b7}, the suggested interval
for $M_{\mathrm{CND}}$ proves
that unperturbed two-body relaxation of the
young stellar disc does not lead to the observed
configuration of the WR/OB stars.

Analogously, we have found the interval for allowed values of
the mass, $M_{\mathrm{c}}$, of the cluster of late-type stars
enclosed within the radius
of the CND: $0\leq M_{\mathrm{c}}\lesssim M_{\bullet}$.
Unlike the CND, this interval is rather wide and contains,
from the observational point of view,
virtually all reasonable values of $M_{\mathrm{c}}$.
We have confirmed that none of the suggested intervals
depend neither upon the shape of the stellar mass function in the
young stellar disc nor the
density
profile of the cluster of late-type stars.

Let us note, however, that a less symmetric,
e.\,g. flattened, cluster of late-type stars could have an
impact on the evolution of the young stellar disc similarly to the
influence of the CND. Nevertheless, since the current observational data
do not provide any indication of such an asymetry, we have not considered
this possibility in our analysis. Moreover, due to the observed
perpendicular mutual orientation of the CWS and the CND, it is likely
that the CND indeed plays a crucial role in the evolution of the young stellar
disc.

Our results further indicate that the angular momenta of
the majority of the WR/OB stars, which are scattered off the CWS
plane, are expected to point to the same hemisphere with respect to
the CND. This feature, together with the mass of the CND, as well as
the mutual orientation of the CND and the CWS,
represent strongly constrained output parameters
of the scenario investigated in this paper.
It would, therefore, be beneficial to concetrate observational
efforts on the corresponding quantities and improve their accuracy in order to
test the suggested hypothesis.

Finally, let us emphasise that the combined effects of two-body relaxation and
differential
precession are efficient enough to transform the
young stellar system into a configuration which matches the
observations in a period of time shorter than the
estimated lifetime of the WR/OB stars. On this account, the presence of the CND
in the system
is not necessarilly required during the whole evolution of the young stellar
disc.
%
%
\section*{Acknowledgments}
J.\,H. gratefully appreciates a scholarship from the Deutscher Akademischer
Austausch Dienst (DAAD) and the hospitality of the
Stellar Populations and Dynamics Group at the AIfA.
This work was supported by the Czech Science Foundation
via grants GACR-205/09/H033 and GACR-202/09/0772
and also from the project SVV~261301
of Charles University in Prague. The calculations were performed on
the computational cluster Tiger at the Astronomical Institute of Charles
University in Prague
({\tt{}http://sirrah.troja.mff.cuni.cz/tiger}).
%
%

%
\bsp
\label{lastpage}
\end{document}